\documentclass[%
 reprint,
 amsmath,amssymb,
 aps,
]{revtex4-1}

\usepackage{graphicx}
\usepackage{dcolumn}
\usepackage{bm}

\usepackage{amsmath}
\usepackage{bbm}

\usepackage{amsmath,amssymb,amsthm,enumerate}

\newcommand{\p}{\partial}

\newcommand{\thetbn}{\arabic{nomer}}

\newcounter{tbn}

{\theoremstyle{definition}

\begin{document}

\preprint{APS/123-QED}

\title{Group properties and invariant solutions of a sixth-order thin film equation\\ in viscous fluid}

%
%
%

\author{Ding-jiang Huang$^{1,2,3}$}
 \altaffiliation {djhuang@fudan.edu.cn}
\author{Qin-min Yang$^{1}$}%
\author{Shuigeng Zhou$^{2,3}$}
\affiliation{%
 $^1$Department of Mathematics, East China University of Science and Technology, Shanghai 200237, China 
}%
\affiliation{
 $^2$School of Computer Science, Fudan University, Shanghai 200433, China
}%
\affiliation{
$^3$Shanghai Key Lab of Intelligent Information Processing, Fudan University, Shanghai 200433, China\\
}%

\date{\today}

\begin{abstract}
Using group theoretical methods, we analyze the generalization of a one-dimensional sixth-order thin film equation which arises in considering the motion of a thin film of viscous fluid driven by an overlying elastic plate. The most general Lie group classification of point symmetries, its Lie algebra, and the equivalence group are obtained. Similar reductions are performed and invariant solutions are constructed. It is found that some similarity solutions are of great physical interest such as sink and source solutions, travelling-wave solutions, waiting-time solutions, and blow-up solutions.
\begin{description}
\item[PACS numbers] 47.15.G-,02.20.Sv,02.30.Jr
\end{description}
\end{abstract}

\pacs{Valid PACS appear here}
\maketitle


\section{\label{sec:level1}Introduction}
In the past several decades there is an increasing interest in physics and mathematics literatures
in higher-order nonlinear diffusion equations because they are models of various interesting phenomenon
in fluid physics and have surprising mathematic structure and properties. Probably, one of the most famous example is
the fourth-order thin film equation in the form
\begin{equation}\label{FTFEq}
u_{t}=(u^{\alpha}u_{xxx})_x,\quad\quad \alpha>0
\end{equation}
which was first introduced by Greenspan in 1978 \cite{Green78}. This equation describes the surface-tension-dominated motion of thin viscous films for the film height $u(t, x)$ and spreading droplets in the lubrication approximation \cite{Green78}. In particular, for $\alpha=3$ it describes a classical thin film of Newtonian fluid, as reviewed in \cite{Oro97}, $\alpha=1$ occurs in the dynamics of a Hele-Shaw cell \cite{Dupont93}
and $\alpha=2$ arises in a study of wetting films with a free contact line between film and substrate \cite{Bert98}. There also exist many interesting generalizations of the famous equation \eqref{FTFEq} (see \cite{King2001,Yarin93} and reference therein).

Apart from the fourth-order equations, another interesting higher-order diffusion model is the sixth-order nonlinear thin film equation in the form
\begin{equation}\label{STFEq}
u_{t}=(u^mu_{xxxxx})_x,
\end{equation}
which appear in flow modeling. The case $m=3$, for instance, was first introduced by King in \cite{King86}  as a model of the oxidation of silicon
in semiconductor devices \cite{King89} or for a moving boundary given by a beam of negligible
mass on a surface of a thin film \cite{Smith96}. Here $u\geq 0$ will be treated as the thickness of a fluid film beneath an elastic plate and
$p=u_{xxxx}$ as the pressure within the film \cite{King86}. The other derivatives of $u$ can in the usual way be assigned different physical meaning, for instance $\Gamma=-u_{xx}$ is the bending moment on the overlying plate and $\Sigma=u_{xxx}$ is the shearing force \cite{Landau86},
here all such expressions are dimensionless. An equation of this type can be used
to model the motion of a thin film of viscous fluid overlain by an elastic plate \cite{King89}; see
also Hobart et al.\ \cite{Hobart00} and Huang et al.\ \cite{Huang02} for possible applications of such modelling
approaches to the wrinkling upon annealing of SiGe films bonded to Si substrates. Other
plausible applications of Eq.\ \eqref{STFEq}, and suitable generalizations thereof, include a simple
model for the influence of a crust on a solidifying melt or for a microfluidic pump (see
Koch et al.\ \cite{Koch97}, for instance).

Eqs.\ \eqref{FTFEq} and \eqref{STFEq} are also the second and the third member of a hierarchy arising from the generalized Reynolds equation
\begin{equation}\label{GenReynoldEq}
u_{t}=(u^mp_{x})_x,
\end{equation}
under different driving forces respectively. For gravity driven flows, we have $p=u$, giving the very
widely studied porous-medium equation (see, for example, Aronson \cite{Aron86}). For surface-tension driven flows we have $p =-u_{xx}$, leading to the fourth-order thin film equation \eqref{FTFEq}. For elastic plate driven flows, we have $p=u_{xxxx}$, which give the sixth-order thin film equation \eqref{STFEq} \cite{King89}.

Up to now, the mathematic structure and properties of the fourth-order thin film equation \eqref{FTFEq} have been widely investigated, including (non-)uniqueness, wetting
behaviour and contact line motion, in particular optimal propagation rates,
waiting time or dead core phenomena and self-similar solutions(see Hulshof \cite{Hul01}, for instance).
Recent years, there are also many researches devoted to symmetry group structure and exact solutions of the fourth-order thin film equations \eqref{FTFEq} and their generalizations\cite{smyth1988high, bernis1991similarity, choudhury1995general, bernis2000dipoles, gandarias2000symmetry, bruzon2003new, qu2006symmetries, gandarias2008equivalence, gandarias2001Medina, Cherniha2010Broadbridge}, or searching for special
invariant finite vector spaces of solutions \cite{Galaktionov2007Exact}.

However, the sixth-order thin film equation \eqref{STFEq} has been much less extensively investigated. It was only a few researches that were devoted to qualitative mathematic properties such as the existence of weak solutions, initial boundary value problems (see Bernis Friedman \cite{bernis1990higher}, King et al \cite{flitton2004moving},
Smith et al.\ \cite{smith1996numerical},  Evans et al.\ \cite{evans2000oxidation}, Barrett et al.\ \cite{barrett2004finite}
for existing studies, the first two being analytical and the others primarily
numerical), while the symmetry group properties and corresponding algebraic structure as well as explicit exact solutions of Eq.\ \eqref{STFEq} still remain open.
Therefore, the aim of the present work is to
find such group properties, algebraic structure and exact solutions.
To do this, we investigate alternatively a more general sixth-order nonlinear diffusion equation in the form
\begin{equation}\label{GenSTFEq}
u_{t}=(f(u)u_{xxxxx})_x,
\end{equation}
than the original equation \eqref{STFEq}, where $f(u)$ is an arbitrary smooth function depending on the geometry of the problem and $f_u\neq0$(
i.e.,~\eqref{GenSTFEq} is a nonlinear equation).

We use the method of Lie groups, one of the powerful tools available to solve nonlinear PDEs, and which was discovered and applied
firstly by S. Lie in the nineteenth century, but only in the last decades has it become a common tool for both mathematicians
and physicists (see for examples \cite{bluman1989symmetries, Ovsiannikov1982group, olver1986applications,Bluman2010applications,Alfinito&Leo1995,Carbonaro1997,Pulov&Uzunov&Chacarov1998,Senthilvelan&Poladian&Sterke2002,Struckmeier&Riedel2002,Stewart2004Momoniat}). The method consists of looking for the infinitesimal generators of a group of point transformations which
leave the equation under study invariant. An important point of the Lie theory is that the conditions for an equation to
admit a group of transformations are represented by a set of linear equations, the so-called ``determining equations",
which are usually completely solvable. Having once found the groups of transformations, one can obtain a number of
interesting results, which include the possibility to reduce a partial differential equation with two independent variables to
an ordinary differential equation with one independent variable, etc.. Solving these reduced equations, one can obtain some particular
solutions for the original equations. These particular solutions are usually called ``similarity solutions" or ``invariant solutions" \cite{bluman1989symmetries, Ovsiannikov1982group, olver1986applications,Bluman2010applications}. When the equation contains ``arbitrary elements"(a variable coefficient deriving from the particular equation of state chosen to characterize the
physical mechanism. ``Arbitrary elements" are functions or variable parameters
whose form is not strictly fixed and can be assigned
freely on the grounds of physical hypotheses about the nature
of the medium under consideration.), the theory gives rise to the problem of group classification of differential equations which is the core stone of modern group analysis \cite{Ovsiannikov1982group,Bluman2010applications}. In particular, in the past several years, a numbers of novel techniques, such as algebraic methods based on subgroup analysis of the equivalence group \cite{Basarab&Zhdanov&Lahno2001,Zhdanov&Lahno1999,Gazeau&Winternitza1992,Popovych&Kunzinger&Eshraghi2010}, compatibility and direct
integration \cite{Ovsiannikov1982group, Ibragimov1994V1} (also referred as the Lie-Ovsiannikov method) as well as their generalizations
(eg.\ method of furcate split \cite{Nikitin&Popovych2001}, additional and conditional
equivalence transformations \cite{Popovych&Ivanova2004NVCDCEs,Huang&Ivanova2007}, extended and
generalized equivalence transformation group, gauging of arbitrary
elements by equivalence transformations \cite{Ivanova&Popovych&Sophocleous2010,Huang&Zhou2011}) have been proposed to solve
group classification problem for numerous nonlinear partial differential equations. Although a great deal of classification was solved by these methods, almost all of them are limited to the equations whose order are lower than four (see \cite{Huang&Zhou2011} for details).

In this paper we extend these new techniques, specific compatibility and direct
integration as well as equivalence transformation techniques, to sixth-order nonlinear diffusion equations.
We first carry out group classification of Eq.\ \eqref{GenSTFEq} under the usual equivalence group.
The Lie group of point symmetries of Eq.\ \eqref{STFEq}, as a special case
of Eq.\ \eqref{GenSTFEq}, and its Lie algebra are also obtained. Then similar reductions of the classification models are performed and
invariant solutions are also constructed. It is found that some
similarity solutions are solutions with physical interest: sink and
source solutions, travelling-wave solutions, waiting-time solutions
and blow-up solutions.

The rest of this paper is organized as follows: In Sec.\ \ref{sec:sc} we derive the equivalence group and perform the group classification related to Eq.\ \eqref{GenSTFEq}. In Sec.\ \ref{sec:sr}, similar reductions of classification models are carried out.
Sec.\ \ref{sec:is} contains examples of some specific exact solutions, including sink and
source solutions, travelling-wave solutions, waiting-time solutions
and blow-up solutions, while in
Sec.\ \ref{sec:cr} some concluding remarks are reported.

\section{\label{sec:sc}SYMMETRY CLASSIFICATION}
Background and procedures of the modern Lie group
theory are well described in literature \cite{bluman1989symmetries, Ovsiannikov1982group, olver1986applications,Bluman2010applications,Popovych&Ivanova2004NVCDCEs,Huang&Ivanova2007}. Without going into the
details of the theory, we present only the results below.

Let
\[ {\bf Q}=\tau(t,x,u)\p_t+\xi(t,x,u)\p_x+\phi(t,x,u)\p_u \]
be a vector field or infinitesimal operator on the space of independent and dependent variables $t,~x,~u$. A
local group of transformations $G$ is a symmetry group of
Eq.\ \eqref{GenSTFEq} if and only if
\begin{equation}\label{InfiCri1}
\rm{pr}^{(6)}{\bf Q}|(\Delta)=0,
\end{equation}
whenever $\Delta=u_t -\big[ f(u) u_{xxxxx}\big]_x=0$ for every generator of $G$,
where $\rm{pr}^{(6)}{\bf Q}$ is the sixth-order prolongation of ${\bf Q}$.

Expanding Eq.\ \eqref{InfiCri1} we get
\begin{eqnarray} \nonumber
\phi^{t} =&& \phi f''(u)u_xu_{xxxxx}+f'(u)u_{xxxxx}\phi^{x}+f'(u)u_x\phi^{xxxxx} \\ &&+\phi f'(u)u_{xxxxxx}+f(u)\phi^{xxxxxx}  \label{InfiCri2}
\end{eqnarray}
which must be satisfied whenever Eq.\ \eqref{GenSTFEq} is satisfied.
Substituting the formulae of $\phi^{t}$, $\phi^{x}$,
$\phi^{xxxxx}$ and $\phi^{xxxxxx}$ into
Eq.\ \eqref{InfiCri2} we get an equation of $t, x, u$ and the
derivatives of $\tau, \xi, \phi, u$. Replacing $u_t$ by the right hand side
of Eq.\ \eqref{GenSTFEq} whenever it occurs, and equating the
coefficients of the various independent monomials to zero, we obtain the determining equations
\begin{equation*}
\begin{cases}
\tau_x=\tau_u=\xi_u=\phi_{uu}=0 \\
3(2\phi_{xu}-5\xi_{xx})f(u)+\phi_xf'(u)=0  \\
(\phi_{xu}-2\xi_{xx})f'(u)=0  \\
3\phi_{xxu}-4\xi_{xxx}=0  \\
(\phi_{xxu}-\xi_{xxx})f'(u)=0  \\
(\tau_t-6\xi_x) f(u)+\phi f'(u)=0 \\
4\phi_{xxxu}-3\xi_{xxxx}=0  \\
\phi_{xxxxxx}f(u)-\phi_t=0 \\
\phi_{xxxxx}f'(u)+\xi_t+6\phi_{xxxxxu}f(u) -\xi_{xxxxxx}f(u)=0  \\
5\phi_{xxxxu}-2\xi_{xxxxx}=0  \\
(5\phi_{xxxxu}-\xi_{xxxxx})f'(u)=0  \\
(2\phi_{xxxu}-\xi_{xxxx})f'(u)=0
\end{cases}
\end{equation*}

The first three equations imply that $\phi_{xu}=\xi_{xx}=0$, which together with
the sixth, the eighth and the ninth equations imply that $\phi_{t}=\phi_{x}=\xi_t=0$,
so the determining equations reduce to
\begin{equation} \label{DEofSTHeq}
\begin{cases}
\tau_x=\tau_u=0 \\
\xi_t=\xi_u=\xi_{xx}=0 \\
\phi_t=\phi_x=\phi_{uu}=0 \\
(\tau_t-6\xi_x) f(u)+\phi f'(u)=0,
\end{cases}
\end{equation}
which is equivalent to
\begin{equation} \label{DE2ofSTHeq}
\begin{cases}
\xi = ax+b \\
\tau = ct+d \\
\phi= pu+q \\
(c-6a)f(u)+(pu+q)f'(u)=0
\end{cases}
\end{equation}
where $a$, $b$, $c$, $d$, $p$, and $q$ are arbitrary constants.

In order to make the classification as simple as possible, we next look for
equivalence transformations of class \eqref{GenSTFEq}, and then solve system \eqref{DE2ofSTHeq} under these transformations.
An equivalence transformation is a nondegenerate change of the variables $t$, $x$ and\ $u$
taking any equation of the form \eqref{GenSTFEq} into an equation of the
same form, generally speaking, with different $f(u)$.
The set of all equivalence transformations forms the
equivalence group $G^{\sim}$. To find the connected component of the unity of\ $G^{\sim}$,
we have to investigate Lie symmetries of the system that consists of
Eq.\ \eqref{GenSTFEq} and some additional conditions, i.e.
\begin{equation} \label{equ1}
\begin{cases}
u_t = f_u u_x u_{xxxxx}+f u_{xxxxxx},  \\
f_t=0, \\
f_x=0.
\end{cases}
\end{equation}
That is to say we
must seek for an operator of the Lie algebra\ $A^{\sim}$ of\ $G^{\sim}$ in the form
\begin{equation}  \label{OperatorEquivTr}
\begin{aligned}
{\bf X}= & \tau(t,x,u)\p_t  + \xi(t,x,u)\p_x \\
         & +\phi(t,x,u)\p_u + \psi(t,x,u,f)\p_f.
\end{aligned}
\end{equation}
Here $u$ and $f$ are
considered as different variables: $u$ is on the space $(t,x)$ and
$f$ is on the extended space $(t, x, u)$. The coordinates
$\tau$, $\xi$, $\phi$ of the operator \eqref{OperatorEquivTr} are sought as functions of
$t$, $x$, $u$ while the coordinates $\psi$ are sought as
functions of $t$, $x$, $u$ and $f$.

Applying $\rm{pr}^{(6)}{\bf X}$ to Eq.\ \eqref{equ1} we get
the infinitesimal criterion
\begin{equation} \label{equ2}
\begin{cases}
\phi^{t}= u_x u_{xxxxx}\psi^u + f_u u_{xxxxx}\phi^{x} \\
\qquad +f_u u_x \phi^{xxxxx}+ u_{xxxxxx} \psi+  f \phi^{xxxxxx},\\
\psi^t=0, \quad \psi^x=0
\end{cases}
\end{equation}
which must be satisfied whenever Eq.\ \eqref{equ1} is satisfied.
Substituting the formulae of $\phi^{t}$, $\phi^{x}$,
$\phi^{xxxxx}$, $\phi^{xxxxxx}$, $\psi^t$, $\psi^x$, and
$\psi^u$ into
Eq.\ \eqref{equ2} we get equations of $t$, $x$, $u$, $f$, and the
partial derivatives of $\tau$, $\xi$, $\phi$, $u$, $f$, and $\psi$.
Replacing $u_t$, $f_t$
and $f_x$ by the right hand side
of Eq.\ \eqref{equ1} whenever they occur, and equating the
coefficients of various
independent monomials to zero,
we obtain
\begin{equation*}
\begin{cases}
\xi_{xx}=0, \quad
\xi_t=0, \quad
\xi_u=0 \\
\tau_x=0, \quad
\tau_u=0 \\
\phi_x=0,\quad
\phi_t=0, \quad
\phi_{uu}=0 \\
\psi_x=0, \quad
\psi_t=0,\quad
\psi_u=0 \\
\psi=(6\xi_x-\tau_t)f
\end{cases}
\end{equation*}
which can be reduced to
\begin{equation} \label{equ3}
\begin{cases}
\tau=c_4t+c_1 \\
\xi=c_5x+c_2 \\
\phi = c_6u+c_3 \\
\psi=(6c_5-c_4)f
\end{cases}
\end{equation}
where $c_1, c_2, \ldots, c_6$ are
arbitrary constants.

Thus the Lie algebra of $G^{\sim}$ for class \eqref{GenSTFEq}
is
\[
A^{\sim}=\langle\p_t, \p_x, \p_u, t\p_t-f\p_f, x\p_x+6f\p_f, u\p_u \rangle.
\]
Continuous equivalence transformations of class \eqref{GenSTFEq} are generated by the
operators from\ $A^{\sim}$. In fact, $G^{\sim}$ contains the following continuous
transformations:
\begin{equation*}
\begin{aligned}
& \tilde t = t {\varepsilon_4}+\varepsilon_1, \qquad
 \tilde x = x  {\varepsilon_5}+\varepsilon_2,   \\
& \tilde u = u {\varepsilon_6} + \varepsilon_3, \qquad
 \tilde f = f \varepsilon_4^{-1}\varepsilon_5^6,
\end{aligned}
\end{equation*}
where $\varepsilon_1, \varepsilon_2, \ldots,\varepsilon_6$ are arbitrary constants.

Solve the last equation of system \eqref{DE2ofSTHeq} under the above equivalence group $G^{\sim}$,
we can obtain three inequivalent equations of class \eqref{GenSTFEq} with respect to the transformations from $G^{\sim}$:

{\bf Case 1:} $f(u)$ is an arbitrary nonconstant function, the symmetry algebra of class \eqref{GenSTFEq} is a three-dimensional Lie algebra which is generated by the operators
\begin{equation} \label{Symm_Opera_arbi}
Q_1=\p_t,\quad
Q_2=\p_x,\quad
Q_3=t \p_t + \frac{1}{6}x\p_x;
\end{equation}

{\bf Case 2:} $f=e^{\lambda u} \mod G^{\sim} (\lambda \neq 0)$, the symmetry algebra of class \eqref{GenSTFEq} is a four-dimensional Lie algebra which is generated by the operators
\begin{equation} \label{Symm_Opera_expo}
\begin{aligned}
&Q_1=\frac{1}{\lambda}\p_u-t\p_t ,
&  Q_2=\p_t, \\
&Q_3=-x\p_x-\frac{6}{\lambda}\p_u,
& Q_4=\p_x;
\end{aligned}
\end{equation}

{\bf Case 3:} $f=u^m \mod G^{\sim} (m \neq 0)$, the symmetry algebra of class \eqref{GenSTFEq} is a four-dimensional Lie algebra which is generated by the operators
\begin{equation} \label{Symm_Opera_powe}
\begin{aligned}
&Q_1=\frac{1}{m}u\p_u-t\p_t ,\quad
&Q_2=\p_t,\\
&Q_3=-x\p_x-\frac{6}{m}u\p_u, \quad
&Q_4=\p_x.
\end{aligned}
\end{equation}

From the above results, it is easy to see that equation \eqref{STFEq} is exactly corresponding to case 3, thus possess a four-dimensional symmetry algebra.

\section{\label{sec:sr}SIMILARITY REDUCTION}

In order to obtain all the inequivalent reductions, we look for the one-dimensional optimal
systems (see \cite{Ovsiannikov1982group}). These systems, similarity variables and reduced equations are listed below.
In the following tables \ref{TableArbitraryFunction}, \ref{TableExponentFunction}, \ref{TablePowerFunction}, each row shows the infinitesimal generators $Q_i$ of each optimal
system, as well as its similarity variable, similarity solution and reduced equation. $\alpha$ is an
arbitrary constant, while $\lambda$ is a non-vanishing arbitrary constant. Note that in the case
$f(u) = u^m$ which corresponds to Eq.\ \eqref{STFEq}, we only consider $m\neq 0$, otherwise the equation is linear.

\subsection{$f(u)$ is an arbitrary nonconstant function}
In this case, the symmetry operators are Eq.\ \eqref{Symm_Opera_arbi}.
These operators satisfy the commutation relations
\[
[Q_1,\ Q_3]=Q_1,\ \ [Q_2,\ Q_3]=\frac{1}{6}Q_2
\]
and thus the corresponding symmetry algebra is a
realization of the algebra $A_{3,5}^a (0<|a|<1)$~\cite{patera1977subalgebras}. According to the results of Patera and Winternitz \cite{patera1977subalgebras},
an optimal system of one-dimensional subalgebras is those spanned by
\[Q_1,\quad Q_2,\quad Q_3,\quad Q_1+\alpha Q_2.\]
Therefore, the corresponding similarity variables and reduced ODEs can be easily calculated. Such results are listed in Table \ref{TableArbitraryFunction}.

\setcounter{tbn}{0}

\begin{center}\footnotesize\renewcommand{\arraystretch}{1.1}
Table \refstepcounter{table}\label{TableArbitraryFunction}\thetable.
Reduced ODEs for arbitrary nonconstant $f(u)$ \\
(let $E=\big[f(v)v_{yyyyy}\big]_{y}$).
\begin{tabular}{|l|c|c|c|l|}
\hline
$i$ & Subalgebra & Ansatz $u=$ & $y$ & \hfil Reduced ODE$_i$ \\
\hline \refstepcounter{tbn}\label{ArbitraryFunction1}\thetbn &
$\langle Q_1\rangle $ & $v(y)$ & $x$
& $E=0$  \\
\refstepcounter{tbn}\label{ArbitraryFunction2}\thetbn &
$\langle Q_2\rangle $ & $v(y)$ & $t$
& $v_{y}=0$  \\
\refstepcounter{tbn}\label{ArbitraryFunction3}\thetbn &
$\langle Q_3\rangle $ & $v(y)$ & $xt^{{-}1/6}$
& $E=-yv_{y}/6$  \\
\refstepcounter{tbn}\label{ArbitraryFunction4}\thetbn &
$\langle Q_1+\alpha Q_2\rangle $ &
$v(y)$ &
$x-\alpha t$ &
$E=-\alpha v_{y}$\\
\hline
\end{tabular}
\end{center}

\subsection{$f(u)=e^{\lambda u}$ ($\lambda \ne 0$)}
In this case, the symmetry operators are given by Eq.\ \eqref{Symm_Opera_expo}, which
satisfy the commutation relations
\begin{equation} \label{com_rela}
[Q_1,\ Q_2]=Q_2,\ \ [Q_3,\ Q_4]=Q_4
\end{equation}
and thus the corresponding symmetry algebra is a
realization of the algebra $2A_{2}$.
According to the results of Patera and Winternitz \cite{patera1977subalgebras} again,
an optimal system of one-dimensional subalgebras is those generated by
\[Q_2, Q_3, Q_4, Q_1+\alpha Q_3, Q_1+\alpha Q_4,
Q_2+\alpha Q_4, Q_2+\alpha Q_3.\]

The corresponding similarity variables and reduced ODEs are listed in Table \ref{TableExponentFunction}.


\begin{center}\footnotesize\renewcommand{\arraystretch}{1.1}
Table \refstepcounter{table}\label{TableExponentFunction}\thetable.
Reduced ODEs for $f(u)=e^{\lambda u}$ \\
(let $\lambda\ne 0$, $E=\big( e^{\lambda v} v_{yyyyy}\big)_{y}$).
\begin{tabular}{|l|c|c|c|l|}
\hline
$i$ & Subalgebra & Ansatz $u=$ & $y$ & \hfil Reduced ODE$_i$ \\
\hline \refstepcounter{tbn}\label{ExponentFunction1}\thetbn &
$\langle Q_2\rangle $ & $v(y)$ & $x$
& $E=0$  \\
\refstepcounter{tbn}\label{ExponentFunction2}\thetbn &
$\langle Q_3\rangle $ & $v(y)+\frac{6}{\lambda}\ln x$ & $t$
& $144e^{\lambda v}-\lambda v_{y}=0$  \\
\refstepcounter{tbn}\label{ExponentFunction3}\thetbn &
$\langle Q_4\rangle $ & $v(y)$ & $t$
& $v_{y}=0$  \\
\refstepcounter{tbn}\label{ExponentFunction4}\thetbn &
$\langle Q_1+\alpha Q_3\rangle $ &
$v(y)+\frac{6\alpha-1}{\lambda}\ln t$ &
$xt^{-\alpha}$ &
$E=\frac{6\alpha-1}{\lambda}-\alpha y v_{y}$\\
\refstepcounter{tbn}\label{ExponentFunction5}\thetbn &
$\langle Q_1+\alpha Q_4\rangle $ &
$v(y)-\frac{1}{\lambda}\ln t$ &
$x+\alpha\ln t$ &
$E=\alpha v_{y}-\frac{1}{\lambda}$\\
\refstepcounter{tbn}\label{ExponentFunction6}\thetbn &
$\langle Q_2+\alpha Q_4\rangle $ &
$v(y)$ &
$x-\alpha t$ &
$E=-\alpha v_{y}$\\
\refstepcounter{tbn}\label{ExponentFunction7}\thetbn &
$\langle Q_2+\alpha Q_3\rangle $ &
$v(y)-\frac{6\alpha t}{\lambda}$ &
$xe^{\alpha t}$ &
$E =\alpha y v_{y}-\frac{6\alpha}{\lambda}$\\
\hline
\end{tabular}
\end{center}

\subsection{$f(u)=u^m$ ($m \ne 0$)}
In this case, the symmetry operators are Eq.\ \eqref{Symm_Opera_powe}.
These operators share the same commutation relations Eq.\ \eqref{com_rela}.
Hence an optimal system of one-dimensional subalgebras is the same as the case
$f(u)=e^{\lambda u}$.
The corresponding similarity variables and reduced ODEs are listed in Table  \ref{TablePowerFunction}.


\begin{center}\footnotesize\renewcommand{\arraystretch}{1.1}
Table \refstepcounter{table}\label{TablePowerFunction}\thetable.
Reduced ODEs for $f(u)=u^m$ \\
(let $m\ne 0$, $E=\big( v^m v_
{yyyyy}\big)_{y}$).
\begin{tabular}{|l|c|c|c|l|}
\hline
$i$ & Subalgebra & Ansatz $u=$ & $y$ & \hfil Reduced ODE$_i$ \\
\hline \refstepcounter{tbn}\label{PowerFunction1}\thetbn &
$\langle Q_2\rangle $ & $v(y)$ & $x$
& $E=0$  \\
\refstepcounter{tbn}\label{PowerFunction2}\thetbn &
$\langle Q_3\rangle $ &
$v(y)x^{\frac{6}{m}}$ & $t$
& $v^{m+1}\prod\limits_{k=-4}^1(\frac{6}{m}{+}k)
=v_{y}$  \\
\refstepcounter{tbn}\label{PowerFunction3}\thetbn &
$\langle Q_4\rangle $ & $v(y)$ & $t$
& $v_{y}=0$  \\
\refstepcounter{tbn}\label{PowerFunction4}\thetbn &
$\langle Q_1+\alpha Q_3\rangle $ &
$v(y)t^{\frac{6\alpha-1}{m}}$ &
$xt^{-\alpha}$ &
$E=\frac{6\alpha-1}{m}v-\alpha y v_{y}$\\
\refstepcounter{tbn}\label{PowerFunction5}\thetbn &
$\langle Q_1+\alpha Q_4\rangle $ &
$v(y)t^{{-}\frac{1}{m}}$ &
$x+\alpha\ln t$ &
$E=\alpha v_{y}-\frac{1}{m} v $\\
\refstepcounter{tbn}\label{PowerFunction6}\thetbn &
$\langle Q_2+\alpha Q_4\rangle $ &
$v(y)$ &
$x-\alpha t$ &
$E=-\alpha v_{y}$\\
\refstepcounter{tbn}\label{PowerFunction7}\thetbn &
$\langle Q_2+\alpha Q_3\rangle $ &
$v(y)e^{-\frac{6\alpha t}{m}}$ &
$xe^{\alpha t}$ &
$E=\alpha y v_{y}-\frac{6\alpha}{m}v$\\
\hline
\end{tabular}
\end{center}

\section{\label{sec:is}INVARIANT SOLUTIONS}
Using the above reduced ODEs, we can construct some invariant solutions for the original equations \eqref{GenSTFEq}. It is easy to see that
some of the similarity variables in the tables \ref{TableArbitraryFunction}, \ref{TableExponentFunction} and \ref{TablePowerFunction} have a clear physical interpretation. Besides, for some higher order reduced ODEs, they can be further reduced by using new symmetries. Below, we
discuss some facts related with some types of similarity solutions with physical interest and obtain some particular solutions. Different types of solutions are separately analyzed.

\subsection{Source and Sink Solutions}
There are two ODEs, i.e., ODE$_{\ref{PowerFunction4}}$ and ODE$_{\ref{PowerFunction7}}$, among our reduced equations are related
to this type of solutions. In fact, if we choose $\alpha=\frac{1}{m+6}$ in ODE$_{\ref{PowerFunction4}}$,
then the similarity solution has the form
\[
u(t,x)=\frac{1}{t^{\frac{1}{m+6}}}v(\frac{x}{t^{\frac{1}{m+6}}}).
\]
Thus, if $m>-6$ it is clear that $u(t,x)\to \delta(x)$ as $t\to 0$
and the similarity solution is a source solution;
if $m<-6$ it is clear that $u(t,x)\to \delta(x)$ as $t\to +\infty$
and the similarity solution is a sink solution. Furthermore, we can also observe that,
for the above choice of $\alpha=\frac{1}{m+6}$, ODE$_{\ref{PowerFunction4}}$
can be integrated once to obtain
\[
v^m v_{yyyyy}+\frac{1}{m+6}yv=k,
\]
where $k$ is an arbitrary constant. Thus, we can obtain a class of source solutions and sink solutions
for the general thin-film equations \eqref{GenSTFEq} with $f(u)=u^m$ (i.e., equation \eqref{STFEq})
by solving the above fifth-order ODE.
If we further choose $k$ as zero, we have
\[v_
{yyyyy}
=-\frac{1}{m+6}yv^{1-m}.
\]
This equation admits the symmetry group corresponding to the infinitesimal generator
${\bf v}= y\p_y+\frac6m v\p_v$.
Taking into account that the invariants of its first prolongation and setting
\[
x_1=v y^{-\frac{6}{m}},\quad u_1=y^{\frac{6}{m}}(yv'-\frac{6}{m}v)^{-1},
\]
this equation becomes
a fourth-order ODE:
\begin{equation*}
\begin{aligned}
&-m^5u_1^3 u_{1x_1x_1x_1x_1}
+5 m^4(3 m u_{1x_1}+ 2 m u_1^2 \\& -6 u_1^2) u_1^2 u_{1x_1x_1x_1}
+10 m^5u_1^2 u_{1x_1x_1}^2
\\&-5 m^3[21 m^2 u_{1x_1}^2 +20 m (m-3) u_1^2 u_{1x_1}\\&+(7 m^2  -48 m+72) u_1^4] u_1 u_{1x_1x_1}
+105 m^5u_{1x_1}^4
\\&+150 m^4 (m-3)u_1^2 u_{1x_1}^3
+15 m^3(7 m^2  -48 m\\&+72) u_1^4 u_{1x_1}^2
+10 m^2 (m-3)(5 m^2-48 m \\& +72) u_1^6 u_{1x_1}
+[m^5 x_1^{-m+1}/(m+6) \\& +72 (m-2) (m-3) (m-6) (2 m-3) x_1] u_1^9 \\&
+12 m(2 m^4-50 m^3+315 m^2\\&-720 m +540) u_1^8=0.
\end{aligned}
\end{equation*}
Thus, source and sink solutions can be also obtained by solving the above fourth-order ODE.

If we choose $m=-6$ in ODE$_{\ref{PowerFunction7}}$,
then the similarity solution has the form
\[u(t,x)=\frac{1}{e^{-\alpha t}}v(\frac{x}{e^{-\alpha t}}),\]
thus, if $\alpha >0$ it is clear that $u(t,x)\to \delta(x)$ as $t\to +\infty$
and the similarity solution is a sink solution;
if $\alpha < 0$ it is clear that $u(t,x)\to \delta(x)$ as $t\to -\infty$
and the similarity solution is a source solution. As in the previous case,
for the choice of $m =-6 $, ODE$_{\ref{PowerFunction7}}$
can be integrated once to obtain a fifth-order ODE
\[
v^{-6} v_{yyyyy}-\alpha y v=k,
\]
where $k$ is an arbitrary constant.
Consequently, source and sink solutions can be computed by solving a fifth-order ODE.

\subsection{Travelling-wave Solutions}
This types of solution corresponds to
the reductions \ref{ArbitraryFunction4}, \ref{ExponentFunction6}
and \ref{PowerFunction6}. In fact, in these three reductions the similarity variables are given
by $y=x-\alpha t$, $u=v$, so that
$u(t,x)=v(x-\alpha t)$, thus the corresponding solutions
are travlling-wave solutions. Due to the physical interest of this type of solutions,
in what follows we study further symmetries of the associated ODEs and then construct
some such kinds of solutions. First of all, we integrate these three equations once
trivially and obtain
\[
\begin{aligned}
\mbox{ODE}'_{\ref{ArbitraryFunction4}}:  f(v)v_{yyyyy}+\alpha v=k,\\
\mbox{ODE}'_{\ref{ExponentFunction6}}:  e^{\lambda v}v_{yyyyy}+\alpha v=k_1,\\
\mbox{ODE}'_{\ref{PowerFunction6}}:  v^mv_{yyyyy}+\alpha v=k_2,
\end{aligned}
\]
where $k, k_1, k_2$ are arbitrary constants. Because ODE$_{\ref{ExponentFunction6}}$ and ODE$_{\ref{PowerFunction6}}$ are given by
ODE$_{\ref{ArbitraryFunction4}}$ for $f(u)=e^{\lambda u}$
and $f(u)=u^m$ respectively, so we will focus on the ODE$'_{\ref{ArbitraryFunction4}}$ below.
This equation is invariant under
the group of translations in the $y$-direction, with infinitesimal
generator $\frac{\p}{\p y}$. Set
\[
x_1=v, \quad u_1=v^{-1}_y,
\]
then the equation becomes a fourth-order ODE:
\begin{equation}  \label{travelling-wave-forth}
  \begin{aligned}
    &f(x_1)(105u^4_{1x_1}-105u_1u^2_{1x_1}u_{1x_1x_1}+10u_1^2u^2_{1x_1x_1}+ \\
    &15u_1^2u_{1x_1}u_{1x_1x_1x_1}
    -u_1^3u_{1x_1x_1x_1x_1})+\alpha x_1u_1^9=ku_1^9.
  \end{aligned}
\end{equation}
We further suppose that
\[
{\bf v} = \xi(x_1,u_1)\p_{x_1}+ \phi(x_1,u_1)\p_{u_1}
\]
is an infinitesimal generator of the last equation, then the coefficients $\xi(x_1,u_1)$ and $\phi(x_1,u_1)$ are satisfied with
\begin{equation} \label{travelling-wave-forth-system}
  \begin{cases}
    \xi = ax_1+b, \\
    \phi = cu_1, \\
    [5 (a+c) \alpha x_1+ b \alpha - ( 4 a + 5 c) k] f(x_1) \\
    +[-a \alpha x_1^2 +( a k - b \alpha) x_1+k b] f'(x_1)=0.
  \end{cases}
\end{equation}

If $f(u)=e^{\lambda u}$, from the above system we can obtain $\xi=\phi=0$, which means that ODE$'_{\ref{ExponentFunction6}}$
has no nontrivial symmetry. Thus, it can not be reduced again. Consequently, the travelling wave solutions for Eq.\ \eqref{GenSTFEq} with $f(u)=e^{\lambda u}$ can be computed by solving a fourth-order ODE:
\begin{equation*} 
  \begin{aligned}
    &e^{\lambda x_1}(105u^4_{1x_1}-105u_1u^2_{1x_1}u_{1x_1x_1}+10u_1^2u^2_{1x_1x_1}+ \\
    &15u_1^2u_{1x_1}u_{1x_1x_1x_1}
    -u_1^3u_{1x_1x_1x_1x_1})+\alpha x_1u_1^9=ku_1^9.
  \end{aligned}
\end{equation*}

If $f(u)=u^m$, then we have three cases from system \eqref{travelling-wave-forth-system}
\[
\begin{array}{ll}
(i)\quad\   k \ne 0, \quad \xi = 0, \quad \phi = 0;\\
(ii) \quad k=0,\quad m=1, \quad \xi = ax_1+b,\quad \phi = -\frac{4}{5}au_1;\\
(iii) \quad  k=0,\quad m\neq 1, \quad \xi = ax_1,\quad \phi = \frac{m-5}{5}au_1.
\end{array}
\]
Due to the triviality, the first case is excluded from the consideration. From the second case, we can get
a rational travelling wave solution for Eq.\ \eqref{GenSTFEq} with $f(u)=u$ in the form
\[
u(t,x)=-\frac{1}{120}\alpha (x-\alpha t)^5+\sum_{i=0}^{4}c_i(x-\alpha t)^i.
\]
For the third case, we can set
\[
x_2=u_1x_1^{\frac{5-m}{5}},\quad u_2=x_1^{\frac{m-5}{5}}(x_1u_1'+\frac{5-m}{5}u_1)^{-1},
\]
then Eq.\ \eqref{travelling-wave-forth} can be reduced to:
\begin{equation*}
\begin{aligned}
&625 x_2^3 u_2^2 u_{2x_2x_2x_2}
-125 [50 x_2 u_{2x_2}+(11 m\\&-25) x_2 u_2^2  +75 u_2] x_2^2 u_2 u_{2x_2x_2}
+9375 x_2^3 u_{2x_2}^3
\\&+125 [3 x_2 u_2 (11 m-  25)+275] x_2^2 u_2 u_{2x_2}^2
\\&+25 [125 (5 m-12) x_2 u_2+(46 m^2  -225 m\\&+250) x_2^2 u_2^2+2625] x_2 u_2^2 u_{2x_2}
+(24 m^4+875 m^2 \\& -250 m^3-1250 m+625 \alpha x_2^5 +625) x_2^4 u_2^7
 \\&+10 (48 m^3 -375 m^2+875 m-625) x_2^3 u_2^6
\\& +125 (38 m^2-195 m +225) x_2^2 u_2^5
+13125 (2 m\\&-5) x_2  u_2^4
 +65625 u_2^3 = 0.
\end{aligned}
\end{equation*}
Consequently, the travelling wave solutions for Eq.\ \eqref{GenSTFEq} with $f(u)=u^m (m\neq 1)$ can be computed by solving a third-order ODE.

Finally, we consider a special situation when $f(u)=ue^{-u}$ and $k=0$, in which system \eqref{travelling-wave-forth-system} infers that ${\bf v} = -5\p_{x_1}+ u_1 \p_{u_1}$.
Let
\[
x_2=u_1 e^{\frac{x_1}{5}}, \quad u_2=-(5u_{1x_1}+u)^{-1}e^{-\frac{x_1}{5}},
\]
then Eq.\ \eqref{travelling-wave-forth} is reduced to:
\begin{equation*}
\begin{aligned}
&x_2^3 u_2^2 u_{2x_2x_2x_2}
-x_2^2 u_2 (10 x_2 u_{2x_2}+11 u_2^2 x_2\\&+15 u_2) u_{2x_2x_2}
+15 x_2^3 u_{2x_2}^3
+11 x_2^2 u_2 (5\\&+3 x_2 u_2) u_{2x_2}^2
+x_2 u_2^2 (46 x_2^2 u_2^2  +125 x_2 u_2\\&+105) u_{2x_2}
+x_2^4 (625 \alpha x_2^5+24) u_2^7
+96 x_2^3 u_2^6  \\&
+190 x_2^2 u_2^5
+210 x_2 u_2^4
+105 u_2^3=0
\end{aligned}
\end{equation*}
Therefore, the travelling wave solutions for equation \eqref{GenSTFEq} with $f(u)=ue^{-u}$ can be computed by solving a third-order ODE too.

\subsection{Waiting-time Solutions}

ODE$_{\ref{PowerFunction2}}$ is a first-order equation that can be easily solved,
in this way we obtain a family of waiting-time solutions for the sixth-order thin film equation \eqref{GenSTFEq} corresponding
to $f(u)=u^m$ (if $m \neq 3/2, 2, 3$ or $6$). These solutions are given by
\[
u(t,x)=\left\{ \begin{aligned}
x^{\frac6m}
\big[m(t_0-t)\prod\limits_{k=-4}^1(\frac6m+k)\big]^{-\frac{1}{m}},\quad x\geq 0, \\
0,\quad\quad\quad\quad\quad\ \ \ \ \ \ \ \ \ \ \ \ \ \ \ \ \ \ \ \ \ \ x<0.
\end{aligned} \right.
\]
where $t_0$ being an arbitrary constant.

\subsection{Blow-up Solutions}

ODE$_{\ref{ExponentFunction2}}$ is also a first-order equation. Solving it we get for
the sixth-order thin film equation \eqref{GenSTFEq} with $f(u)=e^{\lambda u}$ the corresponding similarity solution
\[
u(t,x)=\frac{1}{\lambda}\ln\frac{(x-x_0)^6}{144(t_0-t)}
\]
where $t_0$ is an arbitrary constant. This solution describes a localized blow-up at $x = x_0$. Note that the solution is only valid if
$\frac{(x-x_0)^6}{144(t_0-t)}\leq 1$, then, it ceases before $t = t_0$.
\\

\section{\label{sec:cr}CONCLUDING REMARKS}
We have carried out a detailed group-theoretical analysis for the generalized one-dimensional sixth-order
thin film equation \eqref{GenSTFEq} which arises in considering the motion of a thin film of viscous fluid driven by an overlying elastic plate.
A complete Lie point symmetry group classification for the class ~\eqref{GenSTFEq} have been performed under the continuous equivalence transformation group. Based on these, a complete list of symmetry reductions of the classification cases have been derived by making use of the optimal system of one-dimensional subalgebras of the corresponding Lie symmetry algebras. Furthermore, invariant solutions of the Eq.\ \eqref{GenSTFEq} with different functional form of $f$ have been constructed by solving the reduced ODEs.  In particular, by
focusing our attention in those aspects with physical interest, we have found:

\begin{enumerate}
\item The thin film equation \eqref{GenSTFEq}, for the case $f(u)=u^m$ (which corresponds to equation \eqref{STFEq}), $m>-6$ admits source solutions and $m<-6$ admits sink solutions. These
solutions are related to the solutions of a fourth-order ODE. If $m=-6$, Eq.\ \eqref{GenSTFEq} admits source
and sink solutions. In this case these families of solutions are related to a fifth-order
ODE.

\item The thin film equation \eqref{GenSTFEq} has travelling-wave solutions. In the case $f(u)=u^m$, for $m=1$ the equation admits a
rational travelling-wave solutions, for $m\neq 1$ the problem of finding
these solutions can be transformed into the problem of
solving third-order ODEs.  In the case $f(u)=e^{\lambda u}$, the travelling wave solutions can be computed by solving a fourth-order ODE.
While for the case $f(u)=ue^{-u}$, the travelling wave solutions of equation \eqref{GenSTFEq} can be computed by solving a third-order ODE.

\item Waiting-time solutions in the case $f(u) =u^m$, and blow-up solutions in the case $f(u)=e^{\lambda u}$ are obtained
in the context of symmetry reductions. However, it should be noted that these two
types of solutions can also be obtained by means of variable separation. In the first case
one takes $u(t, x)=T(t)X(x)$ and in the second case $u(t, x) = T(t)+X(x)$.
\end{enumerate}
These results may lead to further applications in physics and engineering such as tests in numerical solutions of Eq.\eqref{GenSTFEq} and as trial
functions for application of variational approach in the
analysis of different perturbed versions of Eq.\eqref{GenSTFEq}. Other topics
including nonclassical symmetry, non-Lie exact solutions and physical applications of class \eqref{GenSTFEq} will be studied
in subsequent publication.

\begin{acknowledgments}
This work was partially supported by the National Key Basic Research Project
of China under Grant No.\ 2010CB126600, the National Natural Science
Foundation of China under Grant No.\ 60873070, Shanghai Leading
Academic Discipline Project No.\ B114, the Postdoctoral Science
Foundation of China under Grant Nos.\ 20090450067, 201104247,, Shanghai
Postdoctoral Science Foundation under Grant No.\ 09R21410600 and
the Fundamental Research Funds for the Central Universities under Grant No.\ WM0911004.
\end{acknowledgments}

\nocite{*}

\bibliography{SixOrderThinFilmEq}

\end{document}